\documentclass[aps,prl,superscriptaddress,showpacs,endfloats,preprint]{revtex4}
\usepackage{graphicx}
\begin{document}
\title{First experimental evidence for quantum echoes in scattering systems}
\author{C.~Dembowski}
\affiliation{Institut f\"{u}r Kernphysik, Technische
Universit\"{a}t Darmstadt, D-64289 Darmstadt, Germany}

\author{B.~Dietz}
\affiliation{Institut f\"{u}r Kernphysik, Technische
Universit\"{a}t Darmstadt, D-64289 Darmstadt, Germany}

\author{T.~Friedrich}
\affiliation{Institut f\"{u}r Kernphysik, Technische
Universit\"{a}t Darmstadt, D-64289 Darmstadt, Germany}

\author{H.-D.~Gr\"{a}f}
\affiliation{Institut f\"{u}r Kernphysik, Technische
Universit\"{a}t Darmstadt, D-64289 Darmstadt, Germany}

\author{A.~Heine}
\affiliation{Institut f\"{u}r Kernphysik, Technische
Universit\"{a}t Darmstadt, D-64289 Darmstadt, Germany}

\author{C.~Mej\'{\i}a-Monasterio}
\affiliation{Center for Nonlinear and Complex Systems, 22100 Como,
Italy}

\author{M.~Miski-Oglu}
\affiliation{Institut f\"{u}r Kernphysik, Technische
Universit\"{a}t Darmstadt, D-64289 Darmstadt, Germany}

\author{A.~Richter}
\affiliation{Institut f\"{u}r Kernphysik, Technische
Universit\"{a}t Darmstadt, D-64289 Darmstadt, Germany}

\author{T.~H.~Seligman}
\affiliation{Centro de Ciencias F\'{\i}sicas, UNAM, Mexico}
\affiliation{Centro Internacional de Ciencias, Cuernavaca, Mexico}
\date{\today}
\begin{abstract}
A self-pulsing effect termed quantum echoes has been observed in
experiments with an open superconducting and a normal conducting
microwave billiard whose geometry provides soft chaos,
\textit{i.e.} a mixed phase space portrait with a large stable
island. For such systems a periodic response to an incoming pulse
has been predicted. Its period has been associated to the degree
of development of a horseshoe describing the topology of the
classical dynamics. The experiments confirm this picture and
reveal the topological information.
\end{abstract}
 \maketitle
Billiards have served as paradigm for chaotic dynamics ever since
the pioneering work of Sinai \cite{sinai}. Flat microwave cavities
of a shape corresponding to billiards have been extensively used
to detect quantum signatures of classical chaos, and served as
analogue systems for properties of atoms, nuclei and molecules
\cite{richter, sridhar, stoeckmannbuch, gmw}. More recently these
analogies became much closer with the discussion of nanostructures
such as quantum dots and quantum wires \cite{imry,stafford}. While
work first concentrated on closed billiards for the study of
spectra and wave functions \cite{richter, stoeckmannbuch}, more
recently interest focussed on scattering systems. On one hand
antennas were considered as open channels
\cite{weidenmueller,stoeckmann1,guhr,schaefer} and on the other
hand outright open billiards, {\it e.g.} billiards connected to
wave guides \cite{smilansky, stoeckmann2,reichl,mendez,luna} have
been studied. In the latter case the wave guides make these
systems rather complicated. Recently there has been a proposition
\cite{jung,jung2,mejia} that open scattering systems with a well
defined finite interaction region may display a self-pulsing
effect called quantum echoes \cite{loschmidtechoes}, if we excite
a system which has a very large stable island in phase space with
a short pulse. In billiards such situations can be reached if the
connection to the exterior region occurs through fairly narrow
necks.  If the classical scattering dynamics is chaotic, we
usually find a chaotic layer which surrounds the stable island in
phase space. The scattering echoes will yield information on the
characteristic structure of this layer \cite{seligman}.
Importantly, this is also true if the observation is of wave
mechanical type. In this sense our experiment is an approach to
the inverse chaotic scattering problem \cite{jung}.

In order to observe the predicted echoes and measure their period,
experiments were performed using flat resonators both at room
temperature and in a superconducting state at $4.2\,\mbox{K}$. The
data were taken for continuous input at fixed frequencies covering
a range from 0 to 20 GHz, and a Fourier transform of these results
yields the response to a short incoming pulse that we require. We
shall first discuss the geometry of the resonator as well as the
location of the antennas that will be used to feed or detect the
microwaves. Next we will consider a surface of section of the
classical motion within this geometry, in order to verify that we
fulfill the conditions to obtain echoes. The relation of the echo
period to the classical scattering dynamics will be touched upon,
and then we proceed to show and discuss the experimental results.

\begin{figure}[ht]
\includegraphics[width=\linewidth]{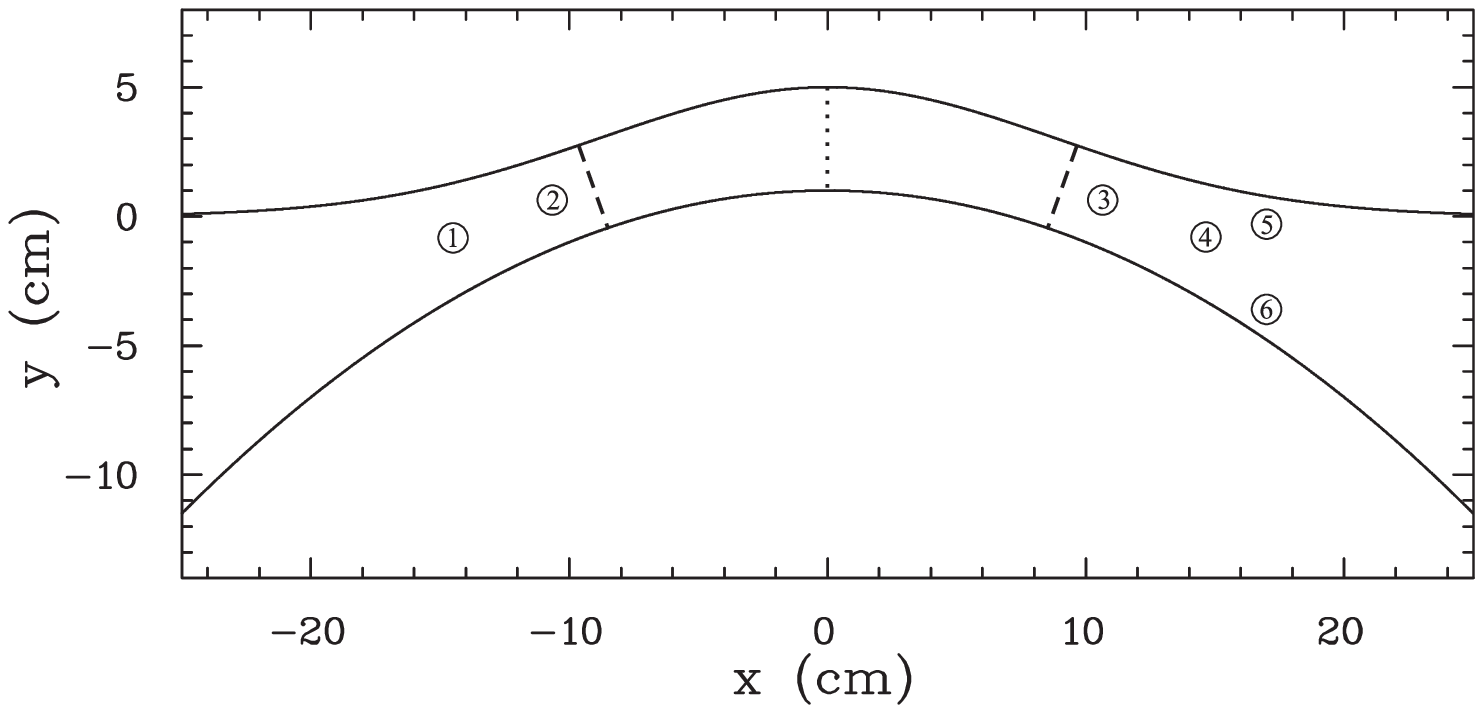}
\caption{The open billiard used in this work possesses a stable
orbit (dotted line) and two unstable periodic orbits (dashed
lines) at the necks. Between the unstable orbits, a wave packet
can get trapped. The circles indicate the positions of the
antennas (labelled by numbers from left to right) used in the
microwave experiment.} \label{f1}
\end{figure}

The geometry of the billiard (Fig.~\ref{f1}) is given by a
Gaussian as an upper boundary and a parabola as a lower one
\begin{equation}
y_u(x)=\lambda\exp{\left(-\frac{\alpha
x^2}{\lambda^2}\right)\mbox{and
}y_l(x)=\lambda\left(\beta-\frac{\gamma x^2}{\lambda^2}\right).}
\label{eq1}
\end{equation}
The parameters have been chosen as $\alpha=0.161$, $\beta=0.2$,
$\gamma=0.1$, and $\lambda=5\,\mbox{cm}$ is a scaling factor. The
billiard extends from $x=-25\,\mbox{cm}$ to $x=25\,\mbox{cm}$ and
shows a stable fundamental periodic orbit along its symmetry axis
at the middle and two unstable ones in the necks where the
distance between both boundaries is minimal. In the following the
part of the billiard between the unstable orbits will be called
the interior, and the complement the exterior. The circles in
Fig.~\ref{f1} refer to the positions of the antennas. They are
located in the exterior of the billiard. Note that a direct
transmission of microwaves through the billiard is not possible.

\begin{figure}[ht]
\includegraphics[width=\linewidth]{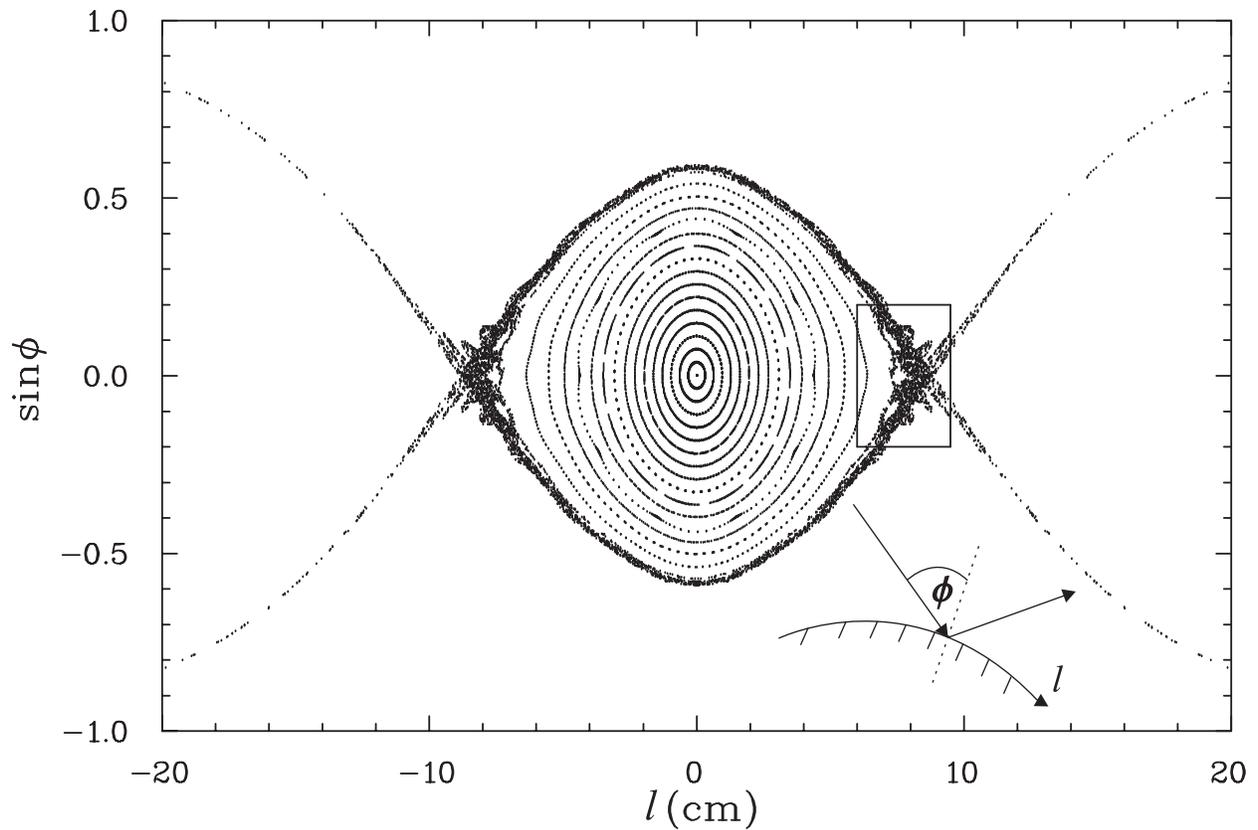}
\caption{A Poincar\'{e} section of the open billiard consists of a
stable island and a surrounding chaotic layer with two hyperbolic
points. Each point in the section corresponds to the angle of
reflection $\phi$ and arc length $l$ of a point of reflection at
the lower boundary of the billiard. The rectangle enclosing the
r.h.s. unstable fixed point shows a heteroclinic structure.}
\label{f2}
\end{figure}

The Poincar\'{e} section of the billiard, which has been obtained
from a classical ray tracing simulation of particles in order to
illustrate the classical trajectories, is shown in Fig.~\ref{f2}.
Each point corresponds to a reflection on the lower boundary,
where the Birkhoff coordinates refer to the arc length of the
point of impact on the boundary and to the tangential component of
the momentum with respect to that boundary. One clearly can see an
island of stability which has been obtained from trajectories
confined to the interior of the billiard. Its center is an
elliptic fixed point corresponding to the fundamental stable
orbit. A chaotic layer which surrounds the island results from the
exterior region, here we find two hyperbolic fixed points,
associated with the two unstable orbits. This situation is
generic. It corresponds to the one of a pendulum with a
perturbation \cite{lichtenberg}. Here the scattering trajectories
will take the place of the revolutions of the pendulum while the
oscillatory island remains in place. Due to the perturbation, the
separatrix is deformed mainly near the two unstable points,
resulting in a chaotic layer.

Particles have a characteristic time for one revolution about the
stable island. Each time they approach one of the necks of the
billiard, they have a chance to escape. Thus from the exterior
regions particles can be observed to leave the interior region
periodically at certain times after an emission of a particle
ensemble, and classical echoes arise.

The rectangular area around the r.h.s. unstable fixed point
displays a complicated structure of tendrils. This is a typical
signature of horseshoe construction (for details, see
\cite{seligman}) underlying the dynamics of the system. This
horseshoe is characterized by a parameter $\beta$, which gives a
hint on the degree of chaoticity of the system and can be obtained
from the lengths and widths of the tendrils. The parameter $\beta$
is related to the period T of the echoes as
\begin{equation}
T=\tau\cdot(-2\log{_3\beta}+3/2)\;. \label{eq2}
\end{equation}
The quantity $\tau$ is the average time between two reflections at
the lower boundary. It can be sufficiently estimated from the size
of the scattering center, i.e. from the length of the stable orbit
in our billiard. The development parameter of the horseshoe can be
expressed as $\beta=3^{-8}$. In this conventional notation, the
base corresponds to the number of outcoupling channels plus one
for the interior region. This yields an echo period of
$T=4.67\,\mbox{ns}$, which matches very well with the result
obtained from the classical simulation of
$T=4.67\pm0.62\,\mbox{ns}$. The error results from the finite
width of the echoes.

The quantum billiard was studied in the experiment using a
microwave cavity sufficiently flat so that the vectorial Helmholtz
equations for the electromagnetic field is reduced to the scalar
Helmholtz equation for the electric field alone which is
equivalent to Schr\"{o}dinger's equation for a particle in a
quantum billiard \cite{richter, stoeckmannbuch}. The billiard was
constructed from lead covered copper plates as in \cite{heine}.

Measurements of the transmission parameters $S_{ij}$ through the
cavity, where $i$ and $j$ denote a pair of antennas, have been
performed for the billiard at room temperature in open air as well
as in the superconducting state when the billiard has been put
into a copper box evacuated and cooled down to $4.2\,\mbox{K}$. In
order to guarantee the openness in the latter case, the openings
of the billiard as well as parts of the box were stuffed with an
urethane based microwave absorber EMC CRAM-AR (HP). By this foam
like material an attenuation of microwave power by two orders of
magnitude was achieved for frequencies larger than
$6\,\mbox{GHz}$. Below this frequency, the absorption is not
uniform, which turned out not to be of disadvantage, because the
measured amplitudes are small in this regime. Furthermore, to show
that the use of these absorbers is equivalent to maintain an open
system, we performed experiments at room temperature in open air
with and without the absorbers. The results coincide.

\begin{figure}[ht]
\includegraphics[width=\linewidth]{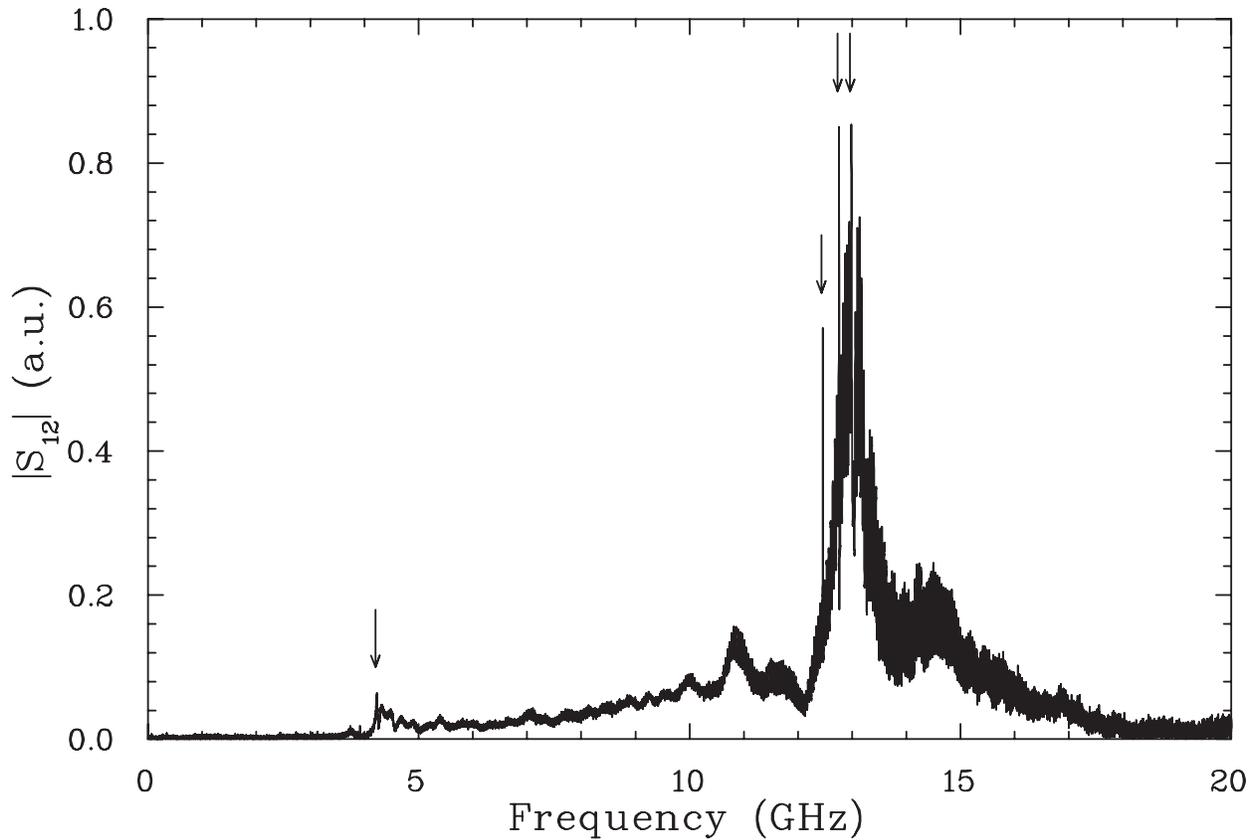}
\caption{Spectrum obtained from a measurement of the transmission
parameter $\vert S_{12}\vert$ between antennas 1 and 2 in the
superconducting resonator. The arrows mark four narrow resonances.
More of those are detected with other antenna combinations. All
sit on top of a continuous spectrum.} \label{f3}
\end{figure}

\begin{figure}[ht]
\includegraphics[width=\linewidth]{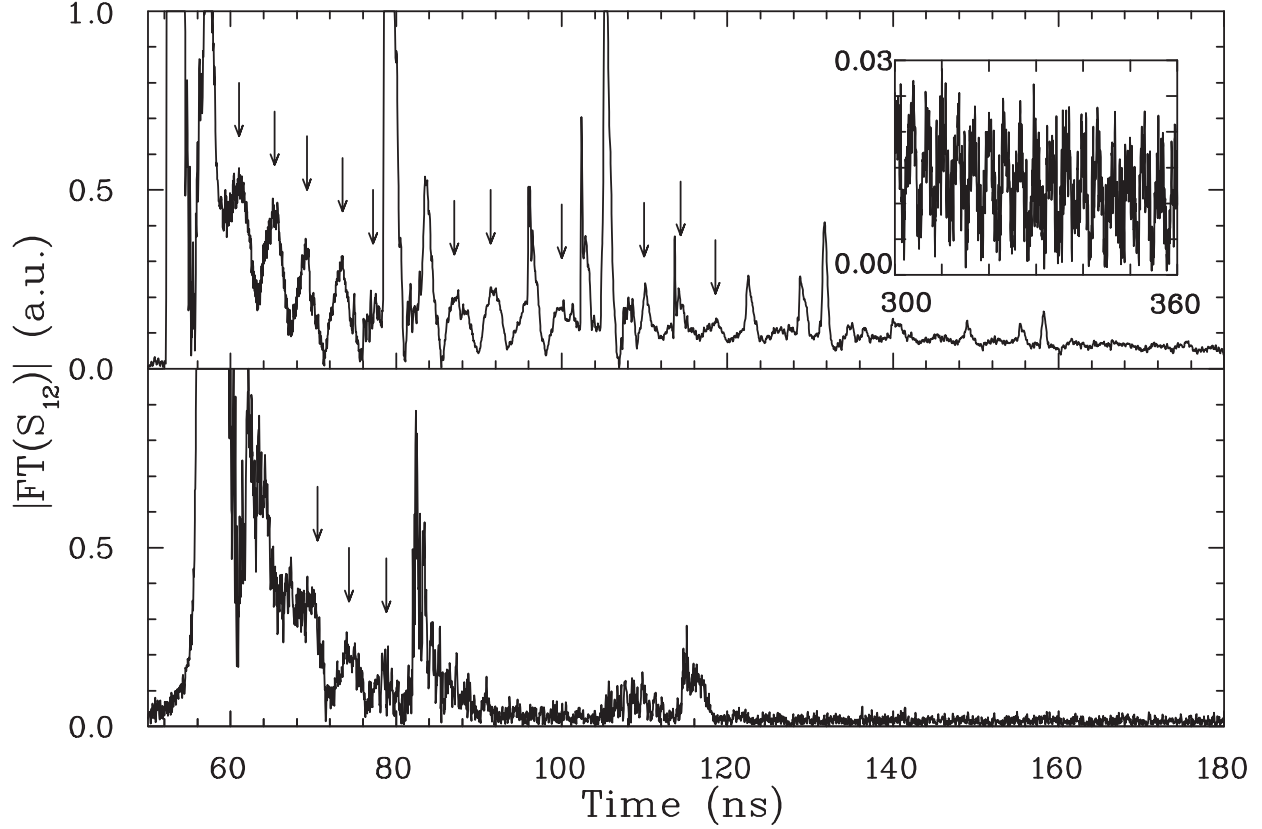}
\caption{Time spectra resulting from a fast Fourier transform of
the spectrum in Fig.~\ref{f3} (upper part) and from the same
measurement at room temperature (lower part). The arrows indicate
the early visible echoes. The additional large peaks are caused by
standing waves on the cables from the RF source to the resonator
and from the resonator to the network analyzer.  The different
offset times in the upper and lower part are due to different
cable lengths. The inset shows echoes at very late times. They
could only be detected in the superconducting case.} \label{f4}
\end{figure}

The measurement consisted of obtaining S-matrix transmission
parameters using a vectorial network analyzer HP-8510C.
Figure~\ref{f3} shows a spectrum for $\vert S_{12}\vert$ in the
superconducting case. The spectrum is continuous, and several
sharp resonances are visible. To get the impulse response in the
time domain, a Fourier transform was performed on all spectra
using a windowed FFT routine. The upper part of Fig.~\ref{f4}
shows the time signals corresponding to the data of Fig.~\ref{f3},
and the inset exhibits the same at much longer times. In the lower
part the signal obtained from the same measurement at room
temperature is displayed. In both time spectra no transmission is
seen below a certain offset time, which reflects the minimum time
the signal needs to propagate through both the cables to and from
the billiard and the billiard itself. One clearly sees periodic
oscillations, which will be shown to be the predicted quantum
echoes. In some measurements with the superconducting resonator,
more than 100 echo periods were identified as can be seen from the
inset. In addition some huge peaks appear in the time spectra.
They are due to standing waves on the cables as has been checked
by varying the cable lengths. Thus they are well understood and do
not affect the experimental results.

\begin{figure}[ht]
\includegraphics[width=\linewidth]{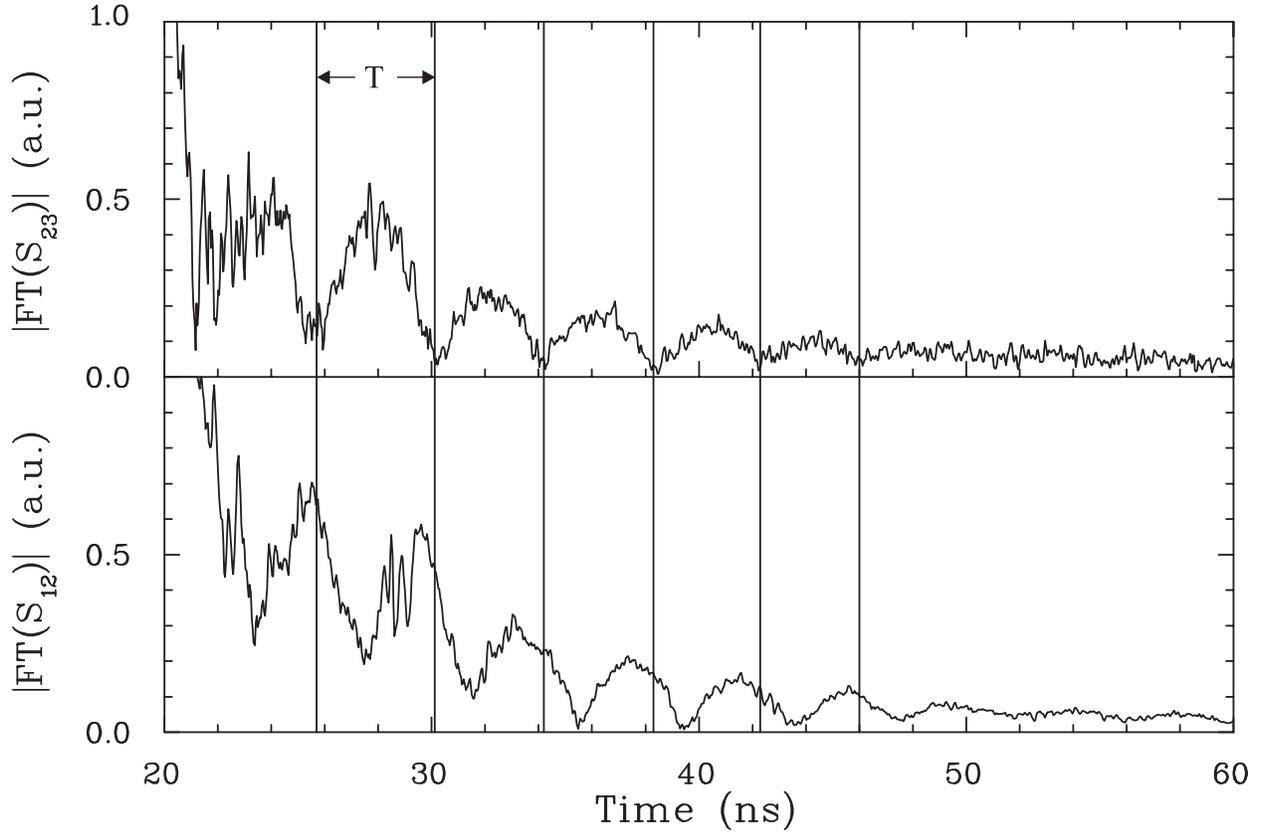}
\caption{Comparison of transmission measurements between antennas
on different sides (upper part) and on the same side (lower part)
of the resonator. A shift of the peaks by about half a period is
visible as expected from the classical model. Note the shortening
of the echo periods $T$ with time.} \label{f5}
\end{figure}

To show that the observed echoes are the ones predicted
\cite{jung}, we first introduced a metal disc into the inner part
of the billiard. From a classical point of view this destroys the
stable island in phase space, which is the theoretical basis for
the echoes. Indeed the experiment reveals, that the echoes
disappear (the corresponding figure is not displayed). Second the
model does not only predict echoes to be detected on both sides of
the cavity outside the necks, but also that those on the r.h.s.
should appear in counter phase to those on the l.h.s. In
Fig.~\ref{f5}, two transmission measurements at room temperature
are presented. The upper part shows the measured $\vert S_{23}
\vert$ parameter between antennas 2 and 3 (c.f. Fig.~\ref {f1}),
while the lower part similarly exhibits $\vert S_{12} \vert $.
Note that the upper part corresponds to a transmission through the
cavity, while the lower part shows a reflection, though measured
by a transmission experiment for two antennas at the same side.
Comparing the upper and lower part of the figure we clearly see
the predicted counter phase behavior.

We shall now discuss two important features of the results. First
it is clear ({\it e.g.} from Fig.~\ref{f5}) that the period of the
echoes gets shorter as a function of time. If we analyze the data
for the superconducting cavity (c.f. Fig.~\ref{f3} upper part) we
find that the period of the echoes starts at
$4.2\pm0.25\,\mbox{ns}$ and slowly decreases to stabilize near
$3.3\pm0.25\,\mbox{ns}$. Second a semi-log plot (not displayed) of
the same data reveals after roughly 10 oscillations an exponential
decay of the average intensity over almost two decades.

Both phenomena can be understood in terms of the classical phase
portrait (c.f. Fig.~\ref{f2}) and the concept of dynamical
tunneling through integrable areas. As our antennas lie outside
the necks no power is injected into the system inside the island.
Thus the inside can exclusively be populated by tunneling, while
for the chaotic layer tunneling may compete with evolution along
classically allowed trajectories. At any rate the intensity will
on average drop with increasing penetration depth into the island.
The time for emission from the chaotic layer and from inside the
island will also increase as we receive the signal from deeper
layers. From the interior of the island the tunneling decay should
be exponential, while we expect both tunneling and direct
contributions from the chaotic layer. It is thus quite clear, that
we receive the outgoing signal, and particularly the echoes from
ever deeper inside the island as time advances. At this point it
is important to note that the rotation period of the island in our
case decreases from the edge, i.e. from the chaotic layer towards
the interior, which is indeed the typical situation, though
exceptions can be constructed. Thus we expect shorter periods for
the echoes as time advances in accordance with the experiment. The
echo period stabilizes and according to this picture we must
assume that we are now seeing tunneling from a fixed penetration
depth of the island. This implies, that the effective barrier is
also fixed and that we should see exponential decay, as we indeed
do. The asymptotic period can either be determined directly by the
innermost states, or more probably by the fact, that as we
penetrate deeper into the island the small absorption of the
cavity wall that remains even in the superconducting case, begins
to dominate the emission through the barrier.

This picture indicates, that the first echoes stem from the edge
of the island or from the chaotic layer, and thus we can invert
Eq.~\ref{eq2} to find a value $-\log{_3\beta}=7.1\pm1.3$. This is
compatible with the value of 8 determined from theory, but
probably even the wave packet causing the first echo penetrates
the island slightly and therefore has a period which is a little
shorter than the classical one.

Summarizing we can conclude the following: We do see the classical
echoes predicted and the properties agree qualitatively and
quantitatively well with what we expect from classical and
semiclassical arguments. It is notable, that this agreement is
achieved in a regime quite far from the semi-classical limit. It
is precisely this fact, that allows the waves to penetrate deep
into the classically forbidden region, and reveal classical
revolution times, that are not accessible in a classical
scattering experiment.

\begin{acknowledgments}
This work was supported by DFG within SFB 634, and by the HMWK
within the HWP. We also acknowledge support from DGAPA-UNAM
project IN101603. B.~D., T.~F., A.~H. and A.~R. acknowledge the
kind hospitality by CIC in Cuernavaca during the workshops
\textit{Billiards: Experiment and Theory} and \textit{Open and
Closed Billiard Systems} in 2003 and 2004.
\end{acknowledgments}

\end{document}